\documentclass[twocolumn,prb, showpacs]{revtex4}% Physical Review B
\usepackage{graphicx}% Include figure files
\usepackage{dcolumn}% Align table columns on decimal point
\usepackage{bm}% bold math
\usepackage{color}

\begin{document}

\title{Superconducting to spin glass state transformation in $\beta$-pyrochlore K$_x$Os$_2$O$_6$}

\author{C. C. Lee$^{1}$}
\author{W. L. Lee$^2$}
\author{J. -Y. Lin$^3$}
\author{C. C. Tsuei$^4$}
\author{J. G. Lin$^1$}
\author{F. C. Chou$^{1,5}$}
\email{fcchou@ntu.edu.tw}
\affiliation{
$^1$Center for Condensed Matter Sciences, National Taiwan University, Taipei 10617, Taiwan}
\affiliation{
$^2$Institute of Physics, Academia Sinica, Taipei 11529, Taiwan}
\affiliation{
$^3$Department of Physics, National Chiao Tung University, HsinChu 30076, Taiwan}
\affiliation{
$^4$IBM, Yorktown Heights, NY 10598, U.S.A.}
\affiliation{
$^5$National Synchrotron Radiation Research Center, HsinChu 30076, Taiwan}

\date{\today}

\begin{abstract}
$\beta$-pyrochore KOs$_2$O$_6$, which shows superconductivity below $\sim$ 9.7K, has been converted into K$_x$Os$_2$O$_6$ (x $\lesssim$ $\frac{2}{3}$-$\frac{1}{2}$) electrochemically to show spin glass-like behavior below $\sim$ 6.1K.  Room temperature sample surface potential versus charge transfer scan indicates that there are at least two two-phase regions for x between 1 and 0.5.  Rattling model of superconductivity for the title compound has been examined using electrochemical potassium de-intercalation.  The significant reduction of superconducting volume fraction due to minor potassium reduction  suggests the importance of defect and phase coherence in the rattling model. Magnetic susceptibility, resistivity, and specific heat measurement results have been compared between the superconducting and spin glass-like samples.

\end{abstract}


\maketitle

\section{\label{sec:level1}Introduction\protect\\ }

%general introduction
Frustration has been one of the most intriguing cooperative phenomena in magnetism.  The geometrically frustrated system becomes one fascinating class of material with emergent novel physical properties by its complex nature, including the layered triangular and Kagome lattice in 2D, and the 3D pyrochlore structure with a corner-sharing tetrahedral sublattice.\cite{Gardner2010, Ramirez1994}   The finding of the first superconducting 3D pyrochlore transition metal oxides Cd$_2$Re$_2$O$_7$ with T$_c$ $\sim$ 1 K in 2001 was surprising.\cite{Hanawa2001}  The following search of superconductivity in the pyrochlore family has raised T$_c$ up to 9.7K in AOs$_2$O$_6$ with A = Cs, Rb, and K since.\cite{Yonezawa2004, Yonezawa2004a, Yonezawa2004b}

% compare A2B2O7 and A2B2O6O'

\begin{figure}
\begin{center}
\includegraphics[width=3.5in]{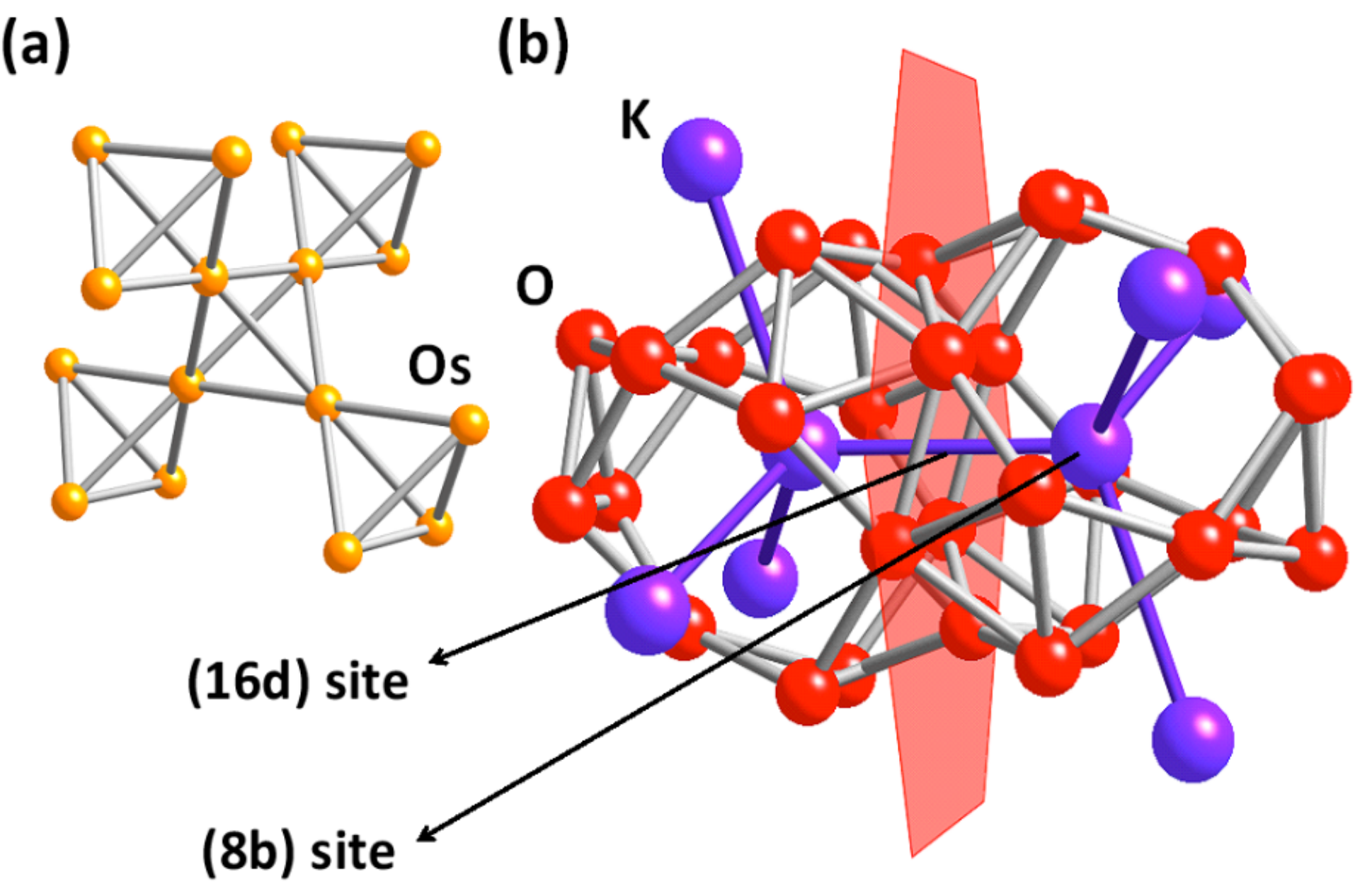}
\end{center}
\caption{\label{fig:fig-structure}(Color online) (a) The osmium ions form 3D frustrated pyrochlore structure. (b) The potassium ions can be viewed as rattling in a cage formed with 18 oxygens, and the cages are connected through tunnels formed with 6 oxygens as indicated by a hexagonal mirror plane.  Ideally K ions sit at 8b site as shown in the center of O$_{18}$ cage, and the center of the O$_6$ pore corresponds to the 16d site. }
\end{figure}

The most general formula of 3D pyrochlore A$_2$B$_2$O$_6$O' is usually described in cubic of Fd$\bar{3}$m (No.227) symmetry with A at 16d ($\frac{1}{2}$,$\frac{1}{2}$,$\frac{1}{2}$), B at 16c (0,0,0), O at 48f (x,$\frac{1}{8}$,$\frac{1}{8}$) and O' at 8b ($\frac{3}{8}$,$\frac{3}{8}$,$\frac{3}{8}$).\cite{Gardner2010}  The 3D pyrochlore A$_2$B$_2$O$_6$O' can be best described as the frustrating conner-shared tetrahedral as shown in Fig~\ref{fig:fig-structure}, where either one or both A and B sublattices can be magnetic.  Different valence combinations including A$^{+3}$/B$^{+3}$, A$^{+2}$/B$^{+4}$, and A$^+$/B$^{+5}$ are possible, although the preferred site of monovalent ion occupies O' site  (8b) instead of A site (16d) as a result of lower and stable electrostatic potential.\cite{Pannetier1973}

The deficient type pyrochlore with monovalent ion occupying 8b (O'-site) instead of 16d (the regular A-site) is called $\beta$-type pyrochlore, in constrast to the regular $\alpha$-type A$_2$B$_2$O$_7$ but written as AB$_2$O$_6$ usually, which is the common structure of the title compound of KOs$_2$O$_6$.   KOs$_2$O$_6$ has a frustrating structure for Os sublattice in corner-shared OsO$_6$ octahedra as shown in Fig.~\ref{fig:fig-structure}(a).  Another aspect to describe the system for K and O relationship is shown in Fig.~\ref{fig:fig-structure}(b), where K is caged by 18 oxygen ions while sitting in the K-tetrahedral center.  As seen in Fig.~\ref{fig:fig-structure}(b), the empty 16d site, which is usually occupied by higher valence A in regular $\alpha$-A$_2$B$_2$O$_7$, is located near the midpoint connecting two Os$_{12}$O$_{18}$ cages within the large pore center formed by 6 oxygens.

%The effort of Na intercalation so far and why: rattling and higher Tc expected
The origin of superconductivity found in (K,Rb,Cs)Os$_2$O$_6$ has been proposed by a rattling model and tested experimentally,\cite{Nagao2009} where effective interaction between two electrons is proposed to be coming from the anharmonic potential generated by the rattling motion of the K$^+$ ions caged in Os$_{12}$O$_{18}$ clusters.  Such rattling-induced superconductivity has been partially supported by the photoemission study as well.\cite{Shimojima2007}  On the other hand, the strong dependence of unit cell size on T$_c$, i.e., the higher T$_c$ from Cs to K family with progressively larger ion size,\cite{Yonezawa2004} seems to suggest that an even higher T$_c$ is waiting to be found if Na substitution becomes possible.  Although Na has been introduced into the system successfully through an ion exchange route from KOs$_2$O$_6$,\cite{Shi2009} the nonstoichiometric product Na$_{1.4}$Os$_2$O$_6$$\cdot$H$_2$O turns out to be non-superconducting down to 2K.  Structure refinement suggests that Na ion does not occupy the original K site at 8b within the Os$_{12}$O$_{18}$ cage, but at the 32e site, which is in the neighborhood of 16d site near the large O$_6$ pore center, i.e., Na$_{1.4}$Os$_2$O$_6$$\cdot$H$_2$O practically belongs to the regular $\alpha$-type pyrochlore instead of  the $\beta$-type pyrochlore such as KOs$_2$O$_6$.

%brief summary of major results
While Na substitution failed to produce the identical structure of KOs$_2$O$_6$ and to raise T$_c$ as expected, we decided to use an alternative route on tuning K content to test the validity of the proposed rattling-induced superconductivity model further. Using the electrochemical chronoamperemetry method, we have reduced potassium content from the pristine KOs$_2$O$_6$ to K$_{x}$Os$_2$O$_6$ (x $\sim$ 0.5) successfully.  We find that the system can be transformed from superconducting to spin glass state below T$_g$ $\sim$6.1K after potassium is reduced to half of its original level.  The resistivity and specific heat measurement results are also reported for a detailed comparative study.

\section{\label{sec:level1}Experiment\protect\\}

Polycrystalline sample KOs$_2$O$_6$ was prepared from reagent-grade oxide powders of KO$_2$ and OsO$_2$ mixed with an appropriate molar ratio (KO$_2$:OsO$_2$ = 1:2) in an argon-filled glove box and pressed into a pellet with mass of $\sim$0.167 g.  The compressed pellet was wrapped with gold foil, sealed in an evacuated glass tube and heated to 450 $^\circ$C for 72 hours.  To control the oxygen partial pressure, a 0.06g Ag$_2$O pellet was added to the sealed glass tube to create an oxygen partial pressure at about $\sim$12 atm at 450 $^\circ$C.   Potassium content of the as-prepared sample is near 1 per formula unit based on the electron probe microanalysis (EPMA) using a spray-on layer of polycrystalline powder, which is consistent with that reported from sample prepared and analyzed using the identical method as before.\cite{Yonezawa2004}

Additional potassium content tuning was accomplished through an electrochemical de-intercalation process.  An electrochemical cell was set up using 1N NaClO$_4$ in propylene carbonate (PC) as electrolyte, Ag/AgCl reference electrode,  and platinum as counter electrode.  Due to the potential formation of hydrated form,\cite{Galati2008} the KOs$_2$O$_6$ sample working electrode was heated to 175 $^\circ$C for 8 hours before the electrochemical experiment and non-aqueous electrolyte was used.  The working electrode was composed of polycrystalline KOs$_2$O$_6$ compressed in pellet form.  An initial chronopotentiometry scan with current density of 1.5 A/g was used to trace the quasi-equilibrium sample potential first.  Final samples were prepared with the chronocoulometry technique, applying voltages from -0.05 to 1.25 V/Ag-AgCl until the induced charge transfer current level had reached time independent constant background.  Potassium content can be reduced progressively  from the original 1 to about $\sim$0.5 by applying constant anodic potentials from -0.05 to 1.25 V relative to the Ag/AgCl Reference electrode.  Crystal structure were analyzed by Bruker D8 X-ray diffractometer.  Magnetic properties were measured using a SQUID magnetometer (Quantum Design SQUID-VSM).  Since powder samples were prepared using a close oxidation environment at a relatively low temperature of 450 $^\circ$C, all following resistivity and specific heat measurement were done on cold compressed pellet under 6 GPa for 20 minutes to reach maximum density.  Resistivity measurements were carried out by a resistance bridge at low current excitation using standard four-probe geometry.  Since the hydration is a reversible process, we made all physical property characterization on samples dried at 175 $^\circ$C for 8 hours beforehand.

\section{\label{sec:level1}Results and Discussions\protect\\ }

\subsection{\label{sec:level2}Electrochemical de-intercalation\protect\\ }

\begin{figure}
\begin{center}
\includegraphics[width=3.5in]{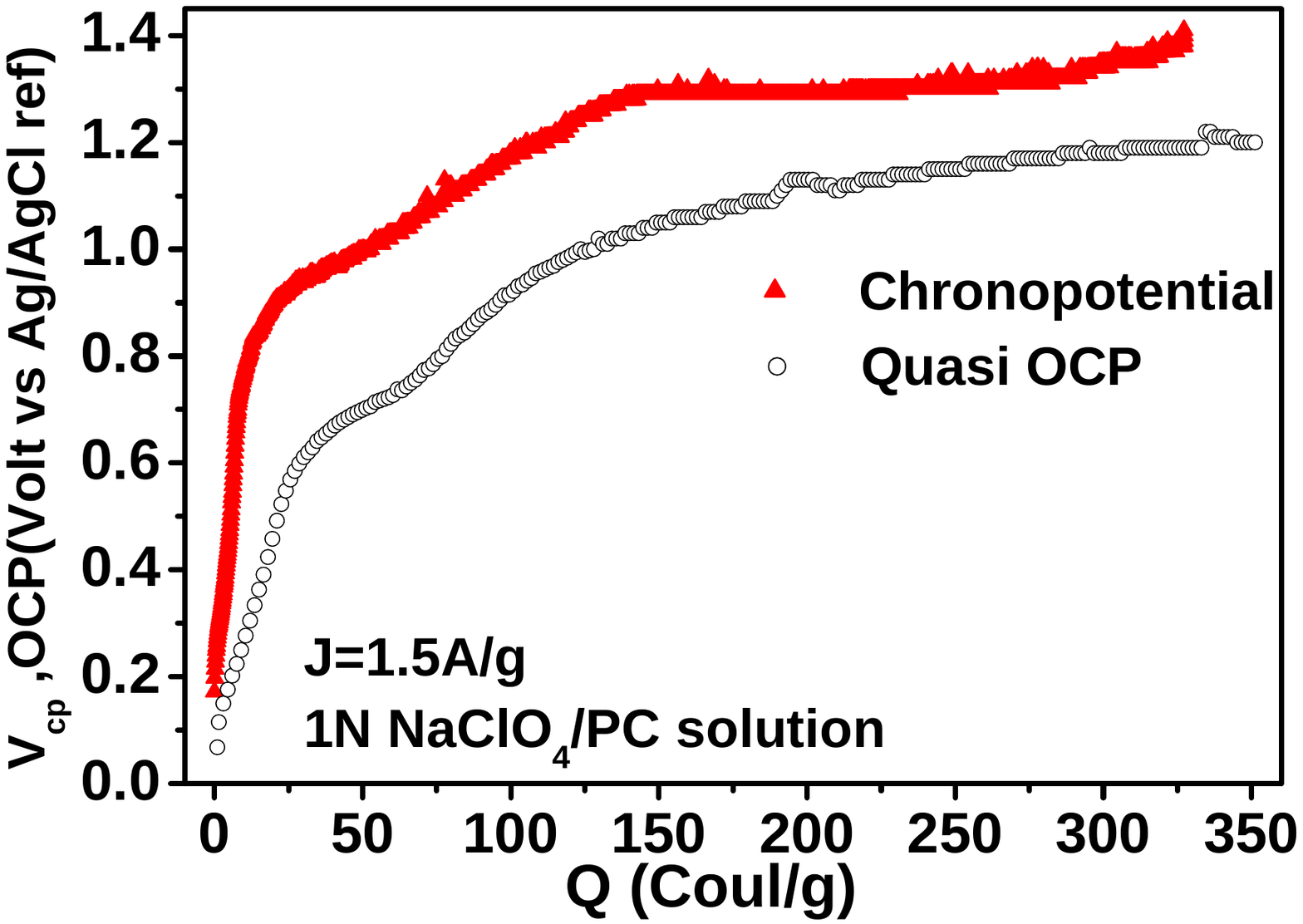}
\end{center}
\caption{\label{fig:fig-xvsV}(Color online) Surface potential and quasi-equilibrium open circuit potential (OCP)versus charge transfer Q for a working electrode made of compressed KOs$_2$O$_6$ powder, where surface potential is obtained from a chronoamparemetry scan of constant current density of 1.5 A/g (red line) and the quasi-equilibrium OCP (black line) is described in the text.  Samples of different potassium content are prepared using applied voltages V$_{ap}$=OCP to produce samples with specific x and verified by EPMA. }
\end{figure}

Since K ions rattle within the large Os$_{12}$O$_{18}$ cage of $\sim$1 $\AA$ anharmonically as suggested by the detectable phonons from fine spectral structure of ARPES experiment,\cite{Shimojima2007} and the ion exchange has been shown to be possible between Na and K ions,\cite{Shi2009} plus the fact that Os ion has 5d electrons of rich valence variation from +2 to +6, we believe that it is highly possible that the reduction of K content can raise the average valence of Os from the current mixed valence of Os$^{+5.5}$ in KOs$_2$O$_6$, to the Os$^{+6}$ state in Os$_2$O$_6$ as a result of the completely emptied cations that potentially occupy the 16d and 8b sites.  To explore the existence of possible stable phases as a function of x in K$_x$Os$_2$O$_6$, we have employed an electrochemical de-intercalation process to control the K level, starting from the as-prepared KOs$_2$O$_6$.  Similar to the reduction of sodium level in Na$_x$CoO$_2$,\cite{Shu2007} potassium de-intercalation becomes possible when ClO$_4$$^-$ is adsorbed to the sample surface and react with the surface K, and the following K ion self diffusion proceeds to reach a new equilibrium at a lower average total K level for the bulk, which should be allowed according to the relatively high electrical and ionic conductivity for the similar pyrochlore structure compounds at room temperature.\cite{Sleight1977}

Chronopotentiometry technique was applied to the system on an electrochemical cell constructed as KOs$_2$O$_6$/1N NaClO$_4$ in PC/Pt.  Non-equilibrium surface potential (V$_{cp}$) versus total charge (Q) passing through sample working electrode is shown in Fig.~\ref{fig:fig-xvsV} under applied constant current density of 1.5 A/g.  The initial open circuit potential V$_{\circ}$ of $\sim$0.1 V has been raised to $\sim$ 0.85 V after nearly one half of the K$^+$ ions had been reduced from the original, assuming 100$\%$ efficiency of charge transfer.  Continued de-intercalation process reduced potassium level further until side reaction was observed after surface potential had reached beyond $\sim$ 1.2 V.  Based on the preliminary chronopotentiometry scan shown in Fig.~\ref{fig:fig-xvsV}, three plateau shape regions corresponding to the two-phase character can be roughly identified, i.e., the first extremely narrow region near V $\sim$ 0.2-0.4, the middle plateau between 0.5-1.1, and the final plateau above $\sim$1.3V.   A significantly higher charge transfer was observed for V$_{cp}$ $\gtrsim$ 1.2V due to side reactions, as observed from the black deposit on the counter electrode Pt surface and volatile vapor bubbling at the cathode.  Since the black deposit found on the Pt counter electrode disappeared after being exposed to the air, it is possible that the deposit was the highly toxic and volatile OsO$_4$.

A separate quasi-equilibrium open circuit potential (OCP) versus Q has been obtained from the same piece of KOs$_2$O$_6$ working electrode \textit{in situ} using chronopotentiometry scan after the current is off for 260 seconds.  Since the background current cannot be estimated accurately before the I(t) curve is fully saturated to the background level, and there may be other undetectable reactions (either anodic or cathodic) included in the total charge transfer, we cannot convert the obtained total charge transfer Q (Coul/g) into actual amount of K de-intercalated accurately, i.e., assuming one electron loss corresponds to one K$^+$ de-intercalated from the KOs$_2$O$_6$ surface.   There are three plateau regions to be noted in the quasi-equilibrated OCP vs. Q diagram, one narrow plateau onset near $\sim$ 0.15V, the second onset near 0.55V, and the final plateau onset near $\sim$ 0.90V which must correspond to the side reaction as discussed above.  There exist at least two distinguishable stable phases that are separated by the two-phase plateau regions: one is proximate to the pristine KOs$_2$O$_6$ phase and the second has stoichiometry close to K$_{0.5}$Os$_2$O$_6$ as later verified by the EPMA chemical analysis.  Following this quasi-equilibrium OCP plot, samples of specific potassium content reported in this study have been prepared using corresponding V$_{ap}$=OCP and final potassium contents were determined based on EPMA chemical analysis.  Samples prepared using different V$_{ap}$ are  summarized in Table~\ref{tab:tableI}.

\begin{figure}
\begin{center}
\includegraphics[width=3.5in]{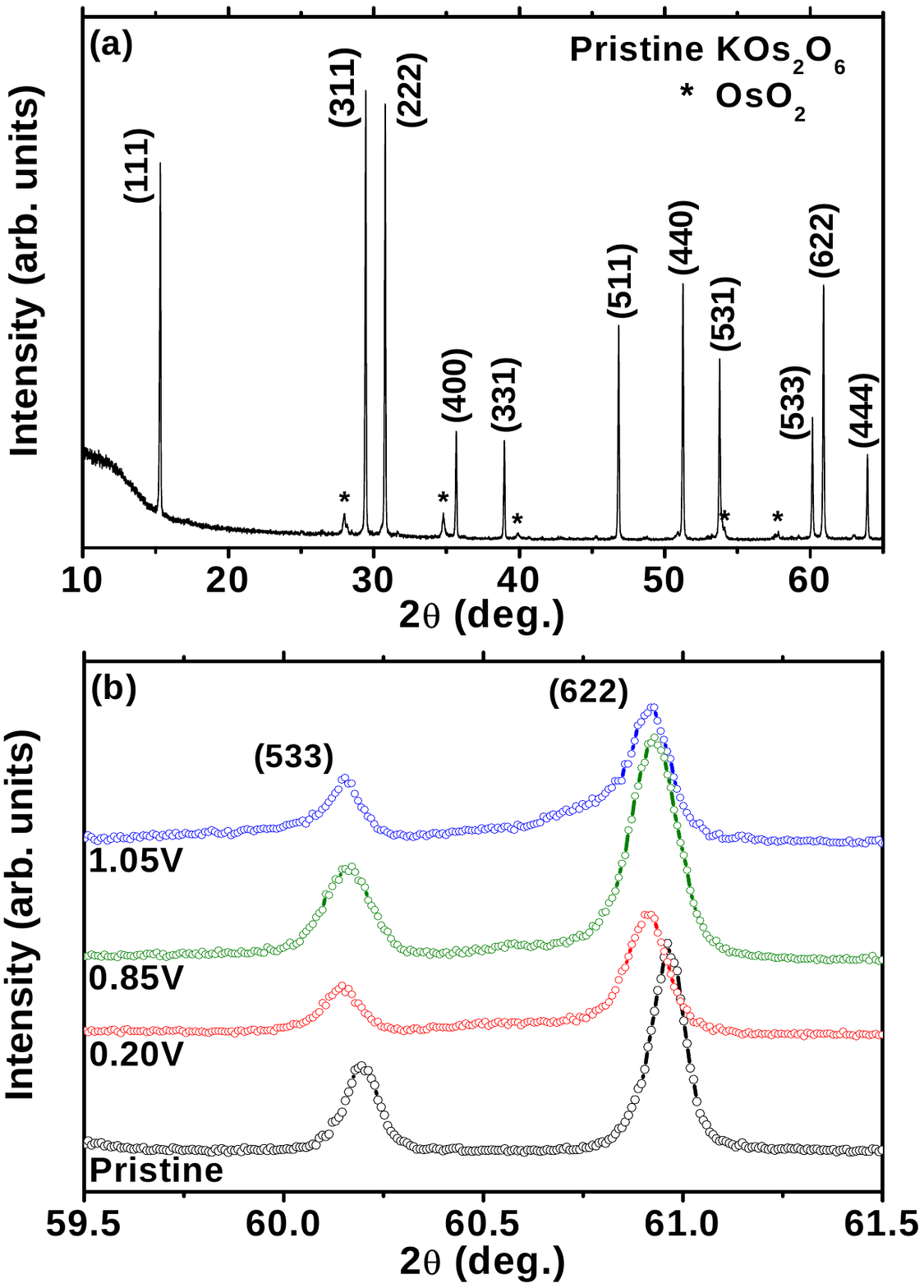}
\end{center}
\caption{\label{fig:fig-xray}(Color online) X-ray diffraction (XRD) patterns for (a) the as-grown KOs$_2$O$_6$, and (b) various de-intercalated K$_x$Os$_2$O$_6$ samples which show slightly shifted diffraction angles.}
\end{figure}

\subsection{\label{sec:level2}X-ray diffraction\protect\\ }

Figure \ref{fig:fig-xray}(a) shows X-ray diffraction (XRD) pattern for the as-grown KOs$_2$O$_6$ powder at room temperature.  All diffraction peaks can be indexed by Fd$\overline{3}$m structure with lattice constant a = 10.073(3) $\AA$.  A few extra peaks labeled with asterisk correspond to the existence of OsO$_2$ impurity phase and it has been reported that the existence of KOsO$_4$ and OsO$_2$ impurity phases are unavoidable for the polycrystalline sample prepared using similar preparation conditions.\cite{Galati2008,Yonezawa2004}   However, we find that the KOsO$_4$ impurity can be washed off with propylene carbonate (PC) solution effectively.  Fig.~\ref{fig:fig-xray}(b) illustrates the XRD patterns at high angles for the de-intercalated KOs$_2$O$_6$ treated with electrochemical de-intercalation at various applied voltages, which indicates the occurrence of a slightly enlarged cell size as a result of application using an even minimal overpotential as low as V$_{ap}$$\sim$ 0.15V.  It is obvious that high angle diffraction peaks shift to lower after the electrochemical processing and no observable extra diffractions peaks to suggest extra impurity phases generated otherwise .

%\begin{figure}
%\begin{center}
%\includegraphics[width=3.5in]{fig-latticesize}
%\end{center}
%\caption{\label{fig:fig-latticesize}(Color online) Applied voltage dependence of lattice constants calculated from peaks indexed (533) and (622) with space group Fd-3m}
%\end{figure}

\begin{table*}
\begin{center}
\caption{\label{tab:tableI} Summary of sample preparation and physical properties.}
\begin{tabular}{cccccc}
\hline\hline

Applied voltage(V)  &  Lattice constant({\AA})  &  Magnetic Behavior  &  T$_{c}$/T$_{g}$  &  S.C. vol. fraction  &  x$_{EPMA}$\\
\hline
pristine&10.072&S.C.&9.7K&20.1$\%$&1.00(4)\\
0.16&10.077&S.C.&9.7K&17.4$\%$&*\\
0.20&10.080&S.C.&9.7K&3.9$\%$&1.01(3)\\
\hline
0.24&10.079&S.C.&9.7K&4.9$\%$&*\\
0.50&10.079&S.C.&9.7K&1.2$\%$&0.97(2)\\
0.64&10.081&S.C.&9.7K&5.8$\%$&*\\
\hline
0.85&10.078&Spin glass&4.6K&*&0.65(6)\\
1.05&10.078&Spin glass&6.1K&*&*\\
1.16&10.078&Spin glass&3.3K&*&0.48(6)\\
\hline\hline
\end{tabular}
\end{center}
\end{table*}

The lattice parameters calculated from high angle diffraction peaks have been indexed with Fd$\overline{3}$m space group and summarized in Table~\ref{tab:tableI}.  We find that the lattice size has an insignificant ($\lesssim$0.01 $\AA$) jump from the pristine superconducting sample of KOs$_2$O$_6$ in general and levels off quickly to be nearly independent of V$_{ap}$ after V$_{ap}$$\sim$0.2V and x.  Such change occurs immediately after a small overpotential is applied with V$_{ap}$ as low as $\sim$ 0.2V to induce K reduction, which is rather surprising since the corresponding potassium is reduced by less than 1 $\%$ only as indicated in Fig.~\ref{fig:fig-xvsV}.  Water intercalation is a common phenomenon for KOs$_2$O$_6$ sample, which causes deterioration of superconductivity and is accompanied by a slight lattice size increase.\cite{Galati2008}  In fact, most reported lattice parameters of the pristine KOs$_2$O$_6$ were possibly overestimated as a result of potential water intercalation, whereas our data fall closest to the sample prepared and processed strictly in a dry box.\cite{Galati2008}  Since the hydration is a reversible process, we made all physical property characterization on samples dried at 175 $^\circ$C for 8 hours beforehand.  Contrary to what is expected from potassium reduction through de-intercalation by progressively higher V$_{ap}$, the lattice size increases while it is expected to decrease when more smaller Os$^{6+}$ ions are generated.  It is highly possible that average valence of Os higher than +5.5 may not be preferred in the pyrochlore structure and induces oxygen defect formation, which is partly supported by the fact that most compounds of high average Os valence prefer perovskite type crystal structure, e.g., Ba$_2$CaOsO$_6$ and Ba$_3$LiOs$_2$O$_9$.\cite{Stitzer2003, Yamamura2006}

\subsection{\label{sec:level2}Magnetic susceptibility\protect\\ }

\begin{figure}
\begin{center}
\includegraphics[width=3.5in]{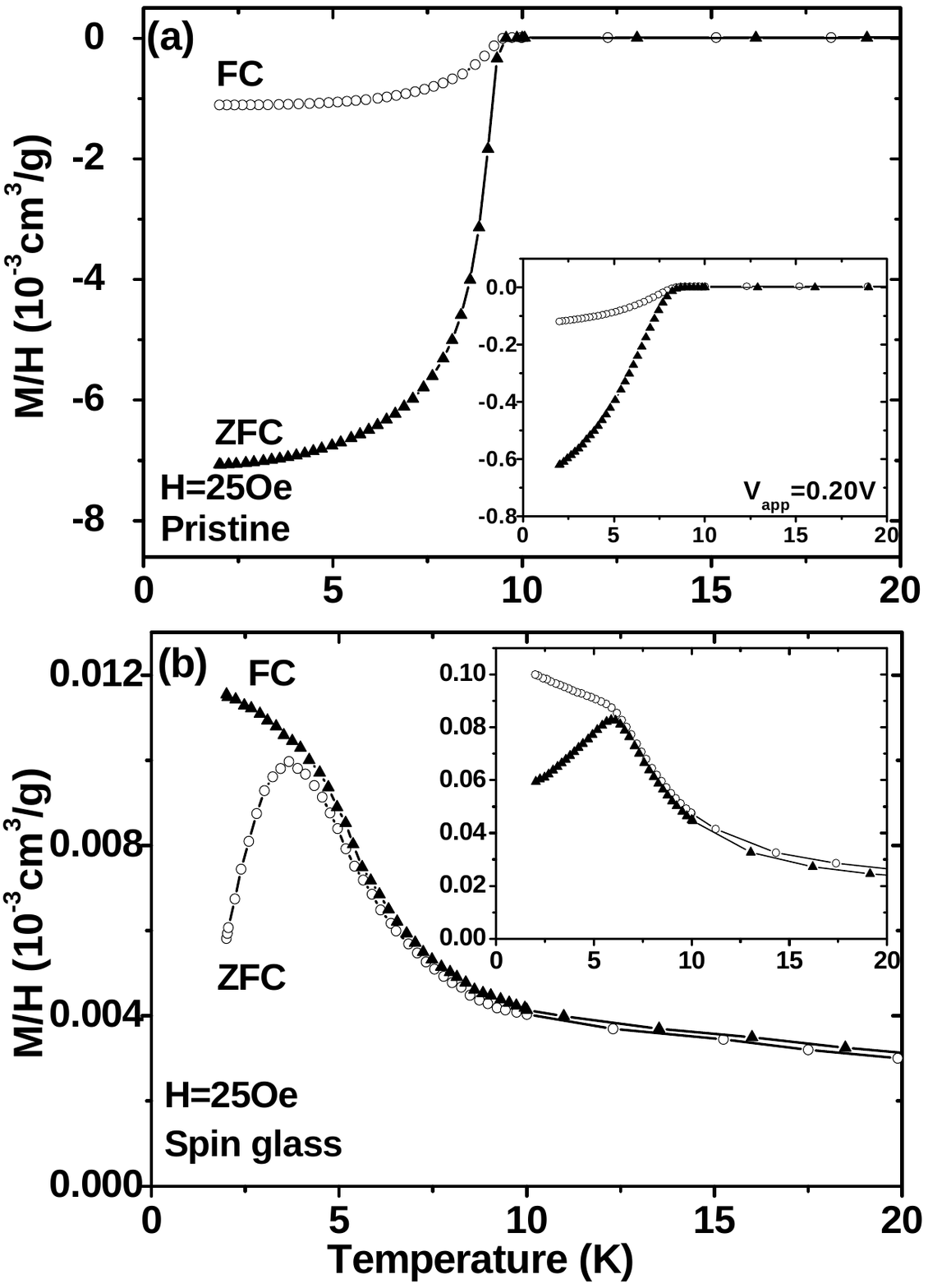}
\end{center}
\caption{\label{fig:fig-chiSC}(Color online) Temperature dependence of dc average susceptibility for K$_x$Os$_2$O$_6$ measured under applied magnetic field of 25 Oe.  (a) shows superconducting sample of the pristine KOs$_2$O$_6$ and the inset shows sample with slightly reduced K using V$_{ap}$=0.20 V, the latter has similar T$_c$ onset but with much lower superconducting volume fraction.  (b) shows spin glass samples of K$_{x}$Os$_2$O$_6$ (x $\lesssim$ 2/3) prepared using V$_{ap}$=1.16 V (T$_g$ $\sim$ 3.3K) and V$_{ap}$=1.05 V (inset, T$_g$ $\sim$ 6.1K), respectively.}
\end{figure}

Bulk superconductivity of T$_c$ $\sim$ 9.7 K has been verified for the pristine KOs$_2$O$_6$ sample as shown in Fig.~\ref{fig:fig-chiSC} by its superconducting volume fraction of $\sim$20$\%$ from the field-cooled data without demagnetization factor correction due to particle geometry.  Interestingly, superconducting volume fraction is significantly reduced after a minor ($\sim$ 1$\%$) potassium reduction using V$_{ap}$ $\gtrsim$ 0.2 V while T$_c$ remains about the same, as shown in the inset of Fig.~\ref{fig:fig-chiSC}(a).  Such drastic reduction is concomitant with the relatively sharp increase in lattice size within the same narrow range below $\lesssim$ 0.20V also shown in Table~\ref{tab:tableI}, beyond which there is no more lattice change and the superconducting volume fraction is reduced to the trace level before falling into the spin glass region.  In fact, a similar drastic change on the surface potential (OCP) has also been observed at the same time for low V$_{ap}$ in the range of $\sim$ 0.16-0.26V (first plateau) as shown in Fig.~\ref{fig:fig-xvsV}.  It is clear that superconductivity exists within an extremely narrow window of stoichiometry KOs$_2$O$_6$ and is sensitive to the lattice size.  From the point of chemical pressure in single band model of conventional BCS picture, the easily enlarged cell size reduces DOS effectively to kill superconductivity.  If we follow the rattling model interpretation, the requirement of highly stoichiometric KOs$_2$O$_6$ implies the importance of phase coherence and intolerance of defects under rattling model, which can be compared with the similar phenomenon of hydration reduced superconductivity, especially when the water molecule occupies the 32e site which is near potassium (8b site) and reduces phonon frequency significantly.\cite{Galati2008}

After passing the two-phase region using V$_{ap}$ $\gtrsim$ 0.85V as indicated by the rough plateau region in Fig.~\ref{fig:fig-xvsV}, single phase sample of K$_{x}$Os$_2$O$_6$ (x$\lesssim$ $\frac{2}{3}$-$\frac{1}{2}$) has been found to show a spin glass-like ZFC/FC irreversible behavior below T$_g$ $\sim$ 6.1 K, as seen in Fig.~\ref{fig:fig-chiSC}(b).  This is the first time observation of spin glass-like phase in the $\beta$-K$_x$Os$_2$O$_6$ system.  Consider the possible random distribution of potassium ions after the reduction, the impact of potassium ion random distribution on its nearby Os spins, and the natural frustration requirement for Os spins in 3D pyrochlore structure, provide the perfect necessary conditions for a canonical spin glass state with localized spins on Os.\cite{Mydosh1993}  While it is difficult to secure reliable structure refinement on the low temperature synthesized powder samples, it is reasonable to assume that the randomness could be coming from the increased occupancy of 16d site (i.e., the preferred site of high valence A in $\alpha$-A$_2$B$_2$O$_7$) when the original 8b site is becoming half empty.  We find that as the potassium content is reduced progressively under higher V$_{ap}$, there is no significant change on the superconducting transition temperature T$_c$ and spin glass-like transition T$_g$, although there is clear relative superconducting volume fraction change, which indicates that only two single phases co-exist, and is in agreement with the middle V-Q plateau as shown in Fig.~\ref{fig:fig-xvsV}.  In addition, the application of overpotential to the sample electrode could also induce possible potassium diffusion from the original 8b site to the 16d site similar to that of the regular $\alpha$-pyrochlore, which can also partly explain the significant reduction of superconducting volume fraction when only less than $\sim$ 1$\%$ potassium is reduced electrochemically as shown in Table~\ref{tab:tableI}.

\begin{figure}
\begin{center}
\includegraphics[width=3.5in]{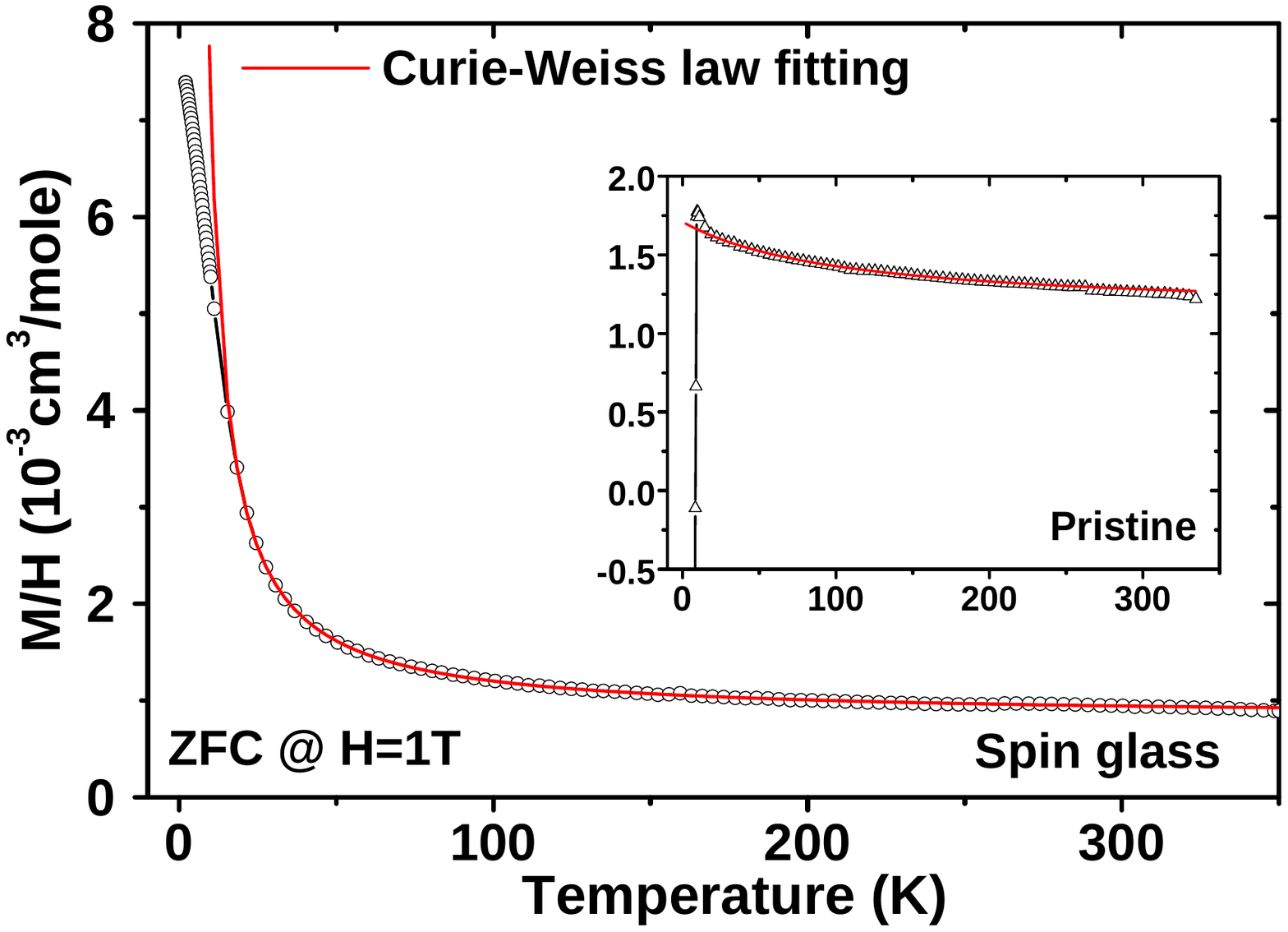}
\end{center}
\caption{\label{fig:fig-chinormal}(Color online) Normal state magnetic susceptibility of K$_x$Os$_2$O$_6$ as a function of temperature under applied field of 1 Tesla for the (a) pristine powder sample and (b) potassium de-intercalated sample using V$_{ap}$ = 1.05 V.  The corresponding low field data are shown in Fig.~\ref{fig:fig-chiSC}.}
\end{figure}

High field magnetic susceptibilities for both the pristine superconducting KOs$_2$O$_6$ and the non-superconducting de-intercalated K$_{0.5}$Os$_2$O$_6$ are shown in Fig.~\ref{fig:fig-chinormal}.  Since the pristine sample is metallic above T$_c$, the application of Curie-Weiss law to describe localized spins is not justified for an itinerant system; in fact, the relatively low and weak temperature dependence of $\chi$(T) does suggest that the contribution seems mostly coming from itinerant Pauli paramagnetic spin contribution, assuming all other temperature independent terms from core diamagnetic and Van Vleck paramagnetic contributions are canceled out.  On the other hand, unlike the usually strongly magnetic 3d and 4d elements, the more spreading 5d orbitals make the electron correlations and spin-orbit interactions more complex and unexpected.  For example, the semiconducting-metal transition near 225 K found in Cd$_2$Os$_2$O$_7$ pyrochlore has been explained in Slater mechanism of local band antiferromagnetic ordering.\cite{Mandrus2001}

Enforcing a Curie-Weiss law fitting for $\chi$(T) as $\chi_{\circ}$+C/(T-$\theta$) to both the superconducting and spin glass samples returns satisfactory fitted values of C$\sim$ 0.061 and 0.035 $cm^3 K/mole$ plus $\theta$ $\sim$ -106K and -8K, respectively.  The Curie constants are significantly lower than that calculated from the theoretical value for a completely localized spin with average valence of +5.5 in high-spin state.\cite{Shi2009} The effective moment for the spin glass sample calculated from fitted C value is $\sim$ 0.53 $\mu_B$.  For K$_{0.5}$Os$_2$O$_6$ in a completely localized ionic picture for Os ion with octahedral oxygen ligands, the average Os valence is +5.75, i.e., 25$\%$ Os$^{5+}$+75$\%$ Os$^{6+}$.  Comparing the fitted $\mu_{eff}$ with the theoretical $\mu_{eff}$ estimated from various possible low spin (LS)/high spin (HS) combinations, the most probable case is that both are in the LS state (i.e., 25$\%$ Os$^{5+}$=1/2 and 75$\%$ Os$^{6+}$=0) with a theoretical spin only $\mu_{eff}$ of 0.43 $\mu_B$.  The LS state assumption is in drastic contrast to most of the Os$^{5+}$ compounds with double perovskite structure in HS state, such as  La$_2$NaOsO$_6$,\cite{Gemmill2005} and the triple perovskite Ba$_3$LiOs$_2$O$_9$ of mostly HS Os$^{5.5+}$,\cite{Stitzer2003}  while 5d orbitals are more spreading compared with the usually magnetic 3d/4d compounds, which makes the electron correlation and spin-orbit coupling more complex.  Current spin only LS implication could be misleading still and its impact to the spin glass formation remains unclear.  Further characterization using larger single crystal of improved homogeneity is necessary. 

The Weiss temperature $\theta$ for the pristine KOs$_2$O$_6$ is comparable to that found in Na$_{1.4}$Os$_2$O$_6$$\cdot$H$_2$O,\cite{Shi2009} but $\theta$ is found to be significantly reduced for the spin glass sample K$_{0.5}$Os$_2$O$_6$, i.e., the antiferromagnetic correlation among isolated spins are significantly reduced for the latter.   The generation of a canonical spin glass state requires satisfaction of both frustration and randomness conditions, which has led to the puzzling observation on the finding of spin glass state in pyrochlore Y$_2$Mo$_2$O$_7$ of unclear origin of randomness.\cite{Greedan1986}  Current observation of spin glass-like behavior in K$_{0.5}$Os$_2$O$_6$ implies that the K$^+$ ions (or oxygen defects, if any) must distribute randomly, most probably between 16d and 8b sites, and lead to the randomness on top of the tetrahedrally coordinated frustrating Os spin substructure.  Because of the limitation on sample quantity and powder inhomogeneity, detailed exploration of spin glass properties using ac spin susceptibility measurement should be performed once sizable single crystal sample is available.

\subsection{\label{sec:level2}Resistivity\protect\\ }

\begin{figure}
\begin{center}
\includegraphics[width=3.5in]{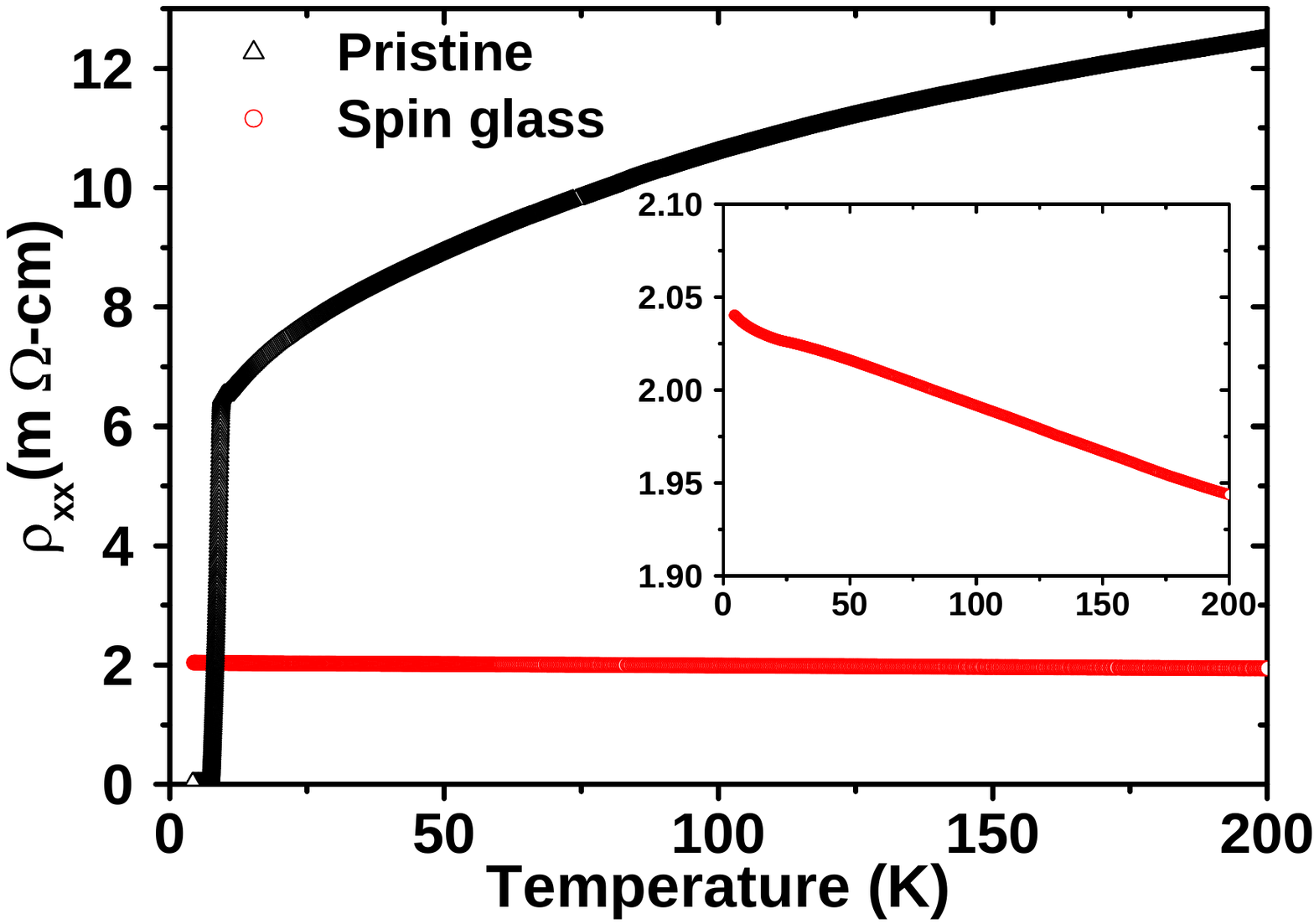}
\end{center}
\caption{\label{fig:fig-resistivity}(Color online) Resistivity versus temperature for the superconducting KOs$_2$O$_6$ and spin glass-like K$_{0.5}$Os$_2$O$_6$ samples.  The spin-glass sample has T$_g$$\sim$3.3K as shown in Fig.~\ref{fig:fig-chiSC}(b).  The inset shows a detailed view for the K$_{0.5}$Os$_2$O$_6$ sample.}
\end{figure}

Resistivity data for both the pristine superconducting sample KOs$_2$O$_6$ and K$_{0.5}$Os$_2$O$_6$ with spin glass-like behavior are shown in Fig.~\ref{fig:fig-resistivity}.  The onset of superconductivity occurs at T$\sim$9.7K for KOs$_2$O$_6$ is confirmed, with $\rho$ monotonically increases with temperature in a concave curvature. This behavior agrees with data previously obtained using single crystal sample,\cite{Dahm2007} where anharmonic potential resulting from the K$^+$ ion rattling plays an important role. However, we do notice that $\rho$(200K)=12.5 m$\Omega$-cm is 50 folds higher than that in a single-crystal. This suggests a significant contribution from the grain boundary scattering in our cold-compressed polycrystalline sample. Nevertheless, the electron-phonon scattering dominates at higher temperature, which gives residual resistivity ratio (RRR) = $\rho$(250K)/$\rho$(10K)$\sim$2. On the other hand, $\rho$(200K) $\sim$ 1.9 m$\Omega$-cm for spin glass-like K$_{0.5}$Os$_2$O$_6$ sample of T$_g$$\sim$3.3K (see Fig.~\ref{fig:fig-chiSC}(b)) is 6 folds smaller than that for the pristine superconducting KOs$_2$O$_6$.  It only increases by 5$\%$ as T drops to 4.2 K as shown in the inset of Fig.~\ref{fig:fig-resistivity}.  The significantly weaker T dependence in $\rho$ for K$_{0.5}$Os$_2$O$_6$ stands in great contrast to that of the superconducting KOs$_2$O$_6$ sample. As pointed out by Nagao \textit{et al.},\cite{Nagao2009} electron-phonon coupling is significantly enhanced by the rattling of alkali ions, which makes KOs$_2$O$_6$ a strong-coupling superconductor with the highest T$_c$ in pyrochlore family.  We therefore attribute the strong suppression of T-dependence of $\rho$ for the spin glass-like K$_{0.5}$Os$_2$O$_6$ is associated with the removal of rattlers (K$^+$ ion), which shuts off the electron-phonon coupling that leads to the superconductivity.  Albeit with the complications from the grain boundary scattering and also the disordering from K$^+$ ion extraction, the weaker T dependence of $\rho$ due to the absence of rattling phonons is further suggested by the specific heat measurements to be discussed next. Further investigation using single crystal sample is needed to clarify this issue.

\subsection{\label{sec:level2}Specific heat\protect\\ }

\begin{figure}
\begin{center}
\includegraphics[width=3.5in]{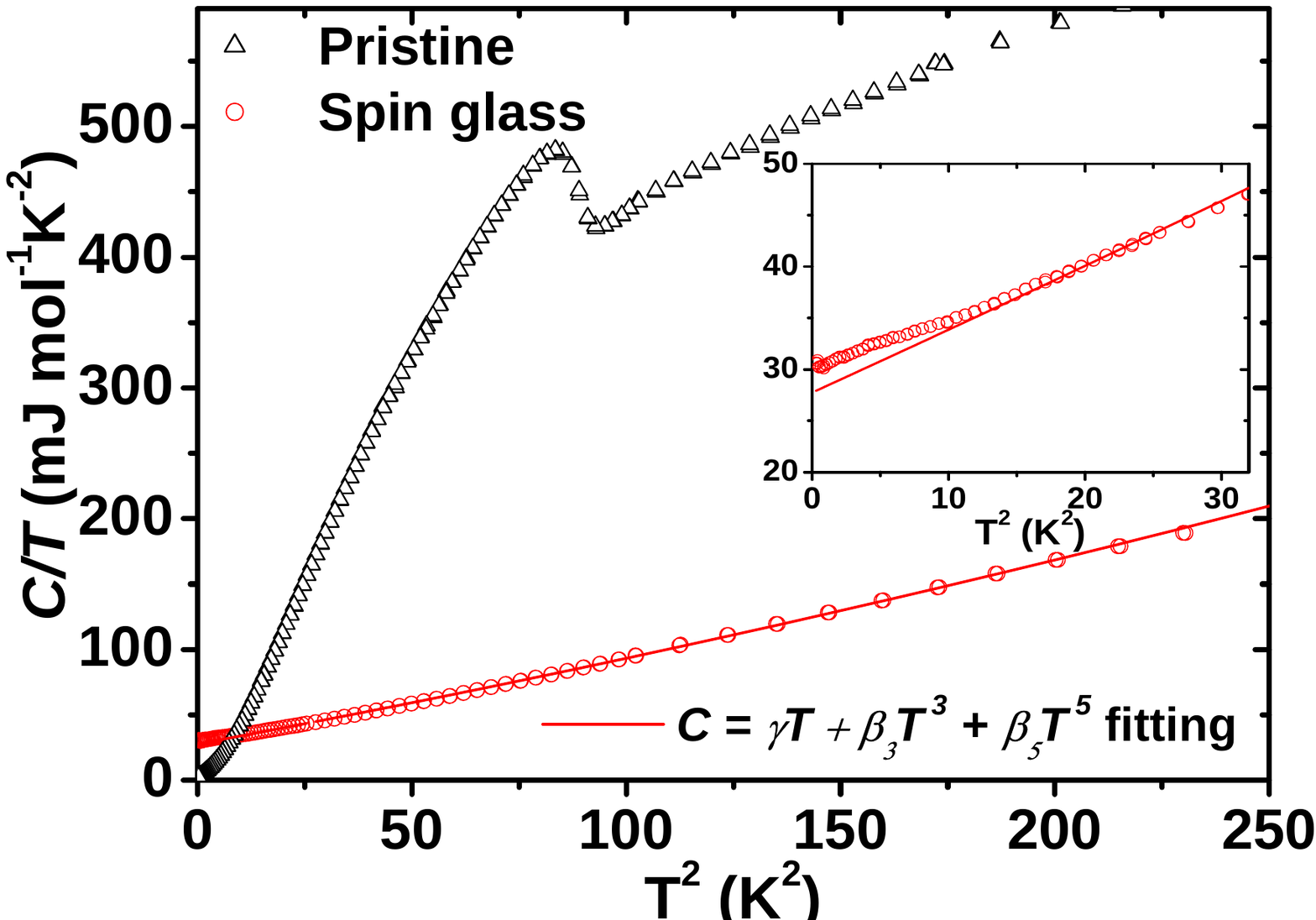}
\end{center}
\caption{\label{fig:fig-Cp}(Color online) Specific heat measurement results of pristine superconducting KOs$_2$O$_6$ and K$_{0.5}$Os$_2$O$_6$.  This particular batch of reduced potassium has a spin glass-like transition of $\sim$ 3.3K as revealed by the ZFC/FC hysteresis onset in magnetic susceptibility as shown in Fig.~\ref{fig:fig-chiSC}(b).}
\end{figure}

Specific heat measurement results for both the pristine and the K$^+$ de-intercalated samples are shown in Fig.~\ref{fig:fig-Cp}.  KOs$_2$O$_6$ has a pronounced anomaly onset $\sim$ 9.7 K due to the superconducting phase transition.  C/T extrapolates to zero at T $\sim$ 0, which indicates a nearly 100$\%$ superconducting volume fraction in KOs$_2$O$_6$.  Hiroi \textit{et al.} have reported the observation of an additional first order phase transition near 7.5K before, which can only be observed in single crystal but never in powder samples.\cite{Yamaura2010, Hiroi2007}  The origin of this additional phase transition below T$_c$ has been suggested to be due to the multipolar transition driven by the octupolar oscillation component.\cite{Hattori2011}  While current specific heat measurement is based on powder sample of greater inhomogeneity comparing with single crystal, it is not surprising for failing to observe such multipolar isomorphic structural transition of first order in nature.     

K$_{0.5}$Os$_2$O$_6$ carries no sign of superconductivity and  behaves like a normal metal. Furthermore, while the magnetization measurements reveal a spin-glass like signature near $\sim$ 3.3K (see Fig.~\ref{fig:fig-chiSC}(b)), there is no apparent associated specific heat peak in this case.  The lack of C(T) peak is exactly one of the features for spin glass phase transition,\cite{Mydosh1993} owing to the much released entropy above T$_g$.  To further illustrate the spin glass signature of C(T) in K$_{0.5}$Os$_2$O$_6$, C(T) is fit by C=$\gamma$T+$\beta_3$T$^3$+$\beta_5$T$^5$ above T$^2$ $>$10 K$^2$, where $\gamma$T and $\beta_3$T$^3$+$\beta_5$T$^5$ correspond to the electronic and phonon contributions.  The fitted results are $\gamma$=27.8 mJ/mol K$^2$, $\beta_3$=0.607 mJ/mol K$^4$ and $\beta_5$=0.00049 mJ/mol K$^6$.    As seen in the inset of Fig.~\ref{fig:fig-Cp}, there exists a small but clear extra contribution beyond the electronic and phonon contributions, which could well be attributed to the spin glass phase.  This spin glass contribution makes C(T) deviate from the C=$\gamma$T+$\beta_3$T$^3$+$\beta_5$T$^5$ at T$^2$ $\lesssim$ 10 K$^2$, in accord with the observed magnetic susceptibility cusp with onset near 3.3 K.

The phonon contribution to C(T) in K$_{0.5}$Os$_2$O$_6$ is another intriguing subject.  In both superconducting samples of CsOs$_2$O$_6$ and RbOs$_2$O$_6$, phonon contributions were described with dominant $\beta_5$T$^5$ term and $\beta_3$ $\sim$ 0.\cite{Nagao2009} This anomalous phonon contribution has been argued to be due to the anharmonicity which led to the alkali ion rattling.  The investigation of the phonon term for KOs$_2$O$_6$ has been claimed to be hindered by its superconductivity with higher T$_c$.  Nevertheless, based on our results shown in Fig.~\ref{fig:fig-Cp}, both the magnitude and temperature dependence of C(T) for KOs$_2$O$_6$ above T$_c$ are very different from those of K$_{0.5}$Os$_2$O$_6$, while the rattling phonon contribution is believed to be similar to that of CsOs$_2$O$_6$ and RbOs$_2$O$_6$.  The Debye temperature derived from $\beta_3$ is $\sim$ 300 K for K$_{0.5}$Os$_2$O$_6$, in contrast to the nearly zero contribution for the reported superconducting samples of (Cs,Rb)Os$_2$O$_6$.\cite{Nagao2009}  The revival of the lattice contribution of the Debye phonons clearly suggests the absence of rattling in K$_{0.5}$Os$_2$O$_6$.  Consequently, the rattling-induced superconductivity is not realized in K$_{0.5}$Os$_2$O$_6$ owing to this drastic change of the electron-phonon interaction, which is consistent with the $\rho$(T) data as shown in Fig.~\ref{fig:fig-resistivity}.  In K$_{0.5}$Os$_2$O$_6$, $\gamma$=27.8 mJ/mol K$^2$ is much smaller than $\gamma$=70 mJ/mol K$^2$ of KOs$_2$O$_6$.  It is not clear whether this smaller $\gamma$ is due to the lower K$^+$ level or the change in band structure owing to the absence of the rattling electron-phonon interaction.

\section{\label{sec:level1}Conclusions\protect\\ }

In summary, spin glass-like behavior below $\sim$ 6.1 K is observed when the potassium level is reduced below $\sim$ $\frac{2}{3}$-$\frac{1}{2}$ from its original level of 1 through an electrochemical de-intercalation process.  The existence of spin glass state has been supported by the thermal hysteresis below T$_g$ in dc spin susceptibility and the deviation from the fitting of specific heat in C(T)=$\gamma$T+$\beta_3$T$^3$+$\beta_5$T$^5$ beyond electronic and phonon contributions below T$_g$.  The randomness and frustration requirement for the existence of spin glass state could be justified by the randomly distributed K$^+$ vacancies and the frustrating tetrahedral coordinated Os sub-structure.    Homogeneous and sizable single crystal sample is desirable for further investigation on the origin of this newly found spin glass phase and its deeper connection to the superconducting state.

Based on the rattling model interpretation of superconductivity, the crossover from the superconducting to spin glass state must be related to the strong suppression of electron-phonon coupling responsible for the occurrence of superconductivity,  which has been supported by both the transport and thermodynamic property measurement results.  While no clear phase diagram has been mapped using the existing powder samples, the proximity of superconducting and spin glass states has been once more demonstrated to be similar to that found in the high T$_c$ and iron pnictide superconducting systems. 

%KOs2O6 with 9.7K superconducting transition temperature is obtained K content can be changed by electrochemical de-intercalation method De-intercalated KxOs2O6 shows larger lattice constant (0.6-0.9ä) Magnetic properties change from superconductor to spin glass for about ³ 8$\%$ K vacancy generates Experimental results indicate that less than 40$\%$ Os+5 ions are with S=1/2 of low spin (LS) state (Assume Os+6 ion with LS=0) For sample with spin state transition form LS of Os+5 or Os+6 to high spin state is possible

\section*{Acknowledgment}
FCC acknowledges the support from National Science Council of Taiwan under project number NSC-98-2119-M-002-021. \\
\appendix

%\bibliography{All-KOs2O6}

\begin{thebibliography}{99}

\bibitem{Gardner2010} J. S. Gardner, M. J. P. Gingras and J. E. Greedan, Rev. Mod. Phys. 82,53 (2010).
\bibitem{Ramirez1994} A. P. Ramirez, Annu. Rev. Mater. Sci. 24, 435 (1994).
\bibitem{Hanawa2001} M. Hanawa, Y. Muraoka, T. Tayama, T. Sakakibara, J. Yamaura, and Z. Hiroi, Phys. Rev. Lett. 87, 187001 (2001).
\bibitem{Yonezawa2004} S. Yonezawa, Y. Muraoka, and Z. Hiroi, J. Phys. Soc. Jpn. 73, 1655 (2004).
\bibitem{Yonezawa2004a} S. Yonezawa, Y. Muraoka, Y. Matsushita, and Z. Hiroi, J. Phys. Soc. Jpn. 73, 819 (2004).
\bibitem{Yonezawa2004b} S. Yonezawa, Y. Muraoka, Y. Matsushita, and Z. Hiroi, J. Phys.: Condens. Matter 16, L9 (2004).
\bibitem{Pannetier1973} J. Pannetier, J. Phys. Chem. Solids 34, 583 (1973).
\bibitem{Nagao2009} Y. Nagao, J.I. Yamaura, H. Ogusu, Y. Okamoto, and Z. Hiroi, J. Phys. Soc. Jpn. 78, 064702 (2009).
\bibitem{Shimojima2007} T. Shimojima, Y. Shibata, K. Ishizaka, T. Kiss, A. Chainani, T. Yokoya, T. Togashi, X.-Y. Wang, C. T. Chen, S. Watanabe, J. Yamaura, S. Yonezawa, Y. Muraoka, Z. Hiroi, T. Saitoh, and S. Shin, Phys. Rev. Lett. 99, 117003 (2007)
\bibitem{Shi2009} Y.G. Shi, A.A. Belik, M. Tachibana, M. Tanaka, Y. Katsuya, K. Kobayashi, K. Yamaura, E. T. Muromachi, J. Solid Stat. Chem. 182, 881 (2009).
\bibitem{Galati2008} R. Galati, C. Simon, C. S. Knee, P. F. Henry, B. D. Rainford, and M. T. Weller: Chem. Mater. 20, 1652 (2008).
\bibitem{Shu2007} G. J. Shu, A. Prodi, S. Y. Chu, Y. S. Lee, H. S. Sheu, and F. C. Chou, Phys. Rev. B 76, 184115 (2007).
\bibitem{Sleight1977} A. W. Sleight, J. E. Gulley, and T. Berzins, Solid Stat. Chem. of Energy Conversion and Storage, Chapter 11, p. 195.
\bibitem{Stitzer2003} K. E. Stitzer, A. E. Abed, M. D. Smith, M. J. Davis, S. J. Kim, J. Darrie,t and H. C. Z. Loye, Inorg. Chem. 42, 947 (2003).
\bibitem{Yamamura2006} K. Yamamura, M. Wakeshima, Y. Hinatsu, J. Solid Stat. Chem. 179, 605 (2006).    
\bibitem{Mandrus2001} D. Mandrus \textit{et al.}, Phys. Rev. B, 63, 195104 (2001).
\bibitem{Greedan1986} J. E. Greedan, M. Sato, and Xu Yan, Solid State Commun. 59, 895 (1986).
\bibitem{Dahm2007} T. Dahm and K. Ueda, Phys. Rev. Lett. 99, 187003 (2007)
\bibitem{Mydosh1993} J. A. Mydosh, Spin glass: an experimental introduction, Taylor $\&$ Francis (1993).
\bibitem{Yamaura2010} J. I. Yamaura, M. Takigawa, O. Yamamuro, and Z. Hiroi, J. Phys. Soc. Jpn. 79, 043601 (2010).
\bibitem{Hiroi2007} Z. Hiroi, S. Yonezawa, and J. I. Yamaura, J. Phys. Condens. Matter 10, 145283 (2007).
\bibitem{Gemmill2005} William R. Gemmill, Mark D. Smith, Ruslan Prozorov, and Hans-Conrad zur Loye, Inorg. Chem., 44, 2639 (2005).
\bibitem{Hattori2011} K. Hattori and H. Tsunetsugu, arXiv:1101.0647v1.

\end{thebibliography}

\end{document}